\documentclass[aps,prl,twocolumn,superscriptaddress,showpacs,letterpaper,reprint]{revtex4-1}

\usepackage{graphicx}
\usepackage{dcolumn}
\usepackage{bm}
\usepackage{amsmath}
\usepackage{multirow}

\begin{document}

\newcommand{\mevcc}{\!\mathrm{MeV}\!/c^2}
\newcommand{\mevc}{\!\mathrm{MeV}/\!c}
\newcommand{\mev}{\!\mathrm{MeV}}
\newcommand{\gevcc}{\!\mathrm{GeV}/\!c^2}
\newcommand{\gevc}{\!\mathrm{GeV}/\!c}
\newcommand{\gev}{\!\mathrm{GeV}}

\title{Electromagnetic Structure of Proton, Pion, and Kaon by High Precision Measurements of their Form Factors at Large Timelike Momentum Transfers}

\author{Kamal~K.~Seth}
\author{S.~Dobbs}
\author{Z.~Metreveli}
\author{A.~Tomaradze}
\author{T.~Xiao}
\affiliation{Northwestern University, Evanston, Illinois 60208, USA}

\author{G. Bonvicini}
\affiliation{Wayne State University, Detroit, Michigan 48202, USA}

\date{22 October 2012}

\begin{abstract} 
The electromagnetic structure of the lightest hadrons, proton, pion, and kaon, is studied by high precision measurements of their form factors for the highest timelike momentum transfers of $|Q^2|=s=14.2$ and 17.4~GeV$^2$. Data taken with the CLEO-c detector at $\sqrt{s}=3.772$~GeV and 4.170~GeV, with integrated luminosities of 805~pb$^{-1}$ and 586~pb$^{-1}$, respectively, have been used to study $e^+e^-$ annihilations into $\pi^+\pi^-$, $K^+K^-$, and $p\bar{p}$.  The dimensional counting rule prediction that at large $Q^2$ the quantity $Q^2F(Q^2)$ for pseudoscalar mesons is nearly constant, and should vary only weakly as the strong coupling constant, $\alpha_S(Q^2)$, is confirmed for both pions and kaons.  
However, the measurements are in strong quantitative disagreement with the predictions of the existing QCD--based models.  For protons, it is found that the timelike form factors continue to remain nearly twice as large as the corresponding spacelike form factors measured in electron elastic scattering, in significant violation of the expectation of their equality at large $Q^2$.  Further, in contrast to pions and kaons, a significant difference is observed between the values of the corresponding quantity $|Q^4|G_M(|Q^2|)/\mu_p$ for protons at $|Q^2|=14.2$~GeV$^2$ and 17.4~GeV$^2$.  The results suggest the constancy of $|Q^2|G_M(|Q^2|)/\mu_p$, instead, at these large $|Q^2|$.
\end{abstract}

\maketitle


Knowledge of the quark--gluon structure of the only stable baryon, the proton, and the lightest mesons, the pion and kaon, is of great interest for both nuclear and particle physics.  Important questions about the size of the proton, the composition of its spin, and the large difference between its spacelike and timelike form factors remain open.  Timelike form factors of pions and kaons, which are needed for the precision determination of the hadronic loop contribution to the muon $g-2$ anomaly~\cite{theory1,theory2}, are poorly known.  Spacelike form factors of pions and kaons needed for the understanding of nuclear and hypernuclear forces are difficult to measure at large momentum transfers, and can only be obtained by analytic continuation of timelike form factors~\cite{miller}.  To meet these needs, precision measurements of timelike form factors at the highest possible momentum transfers are needed.
In this Letter we report measurements of the form factors of pions, kaons, and protons with much higher precision, and for much larger timelike momentum transfers than before~\cite{ffe760-835,ffcleo}.

Earlier measurements of proton form factors for large timelike momentum transfers ($Q^2<0$)  made by the Fermilab E760/E835 $p\bar{p}\to e^+e^-$ experiments for $|Q^2|=8.84-13.11$~GeV$^2$~\cite{ffe760-835}, and  the CLEO $e^+e^-\to p\bar{p}$ measurements at $|Q^2|=13.48$~GeV$^2$~\cite{ffcleo}, revealed that the timelike form factors are nearly twice as large as the corresponding spacelike form factors, a result in strong disagreement with the expectation of their equality at asymptotically large $|Q^2|$.  The measurement of pion and kaon timelike form factors by CLEO at $Q^2=13.48$~GeV$^2$~\cite{ffcleo} revealed that while the dimensional counting rule prediction of a $\alpha_S/|Q^2|$ variation of the form factors~\cite{brodsky} was apparently confirmed, the measured form factors were factors $4-8$ larger than predicted.  Further, the ratio $F_\pi(|Q^2|)/F_K(|Q^2|)$ was also found to be nearly twice as large as the QCD prediction that it should be equal to the ratio of the squares of the decay constants, $f_\pi^2/f_K^2$ at large $Q^2$~\cite{qcd}.  These large differences from theoretical predictions raise important questions about how valid the asymptotic predictions are at the measured momentum transfers, and make it imperative to extend the measurements to larger momentum transfers.  

  We use data taken with the CLEO-c detector,  which has been described in detail before~\cite{cleodetector},  to study the reactions $e^+e^-\to p\bar{p},~\pi^+\pi^-$, and $K^+K^-$.  The main body of the data comprise of $e^+e^-$ annihilations at center-of-mass energies $\sqrt{s}=3772$~MeV ($|Q^2|=14.2$~GeV$^2$) and 4170~MeV ($|Q^2|=17.4$~GeV$^2$), with integrated luminosities of $\mathcal{L}=805$~pb$^{-1}$  and $\mathcal{L}=586$~pb$^{-1}$, respectively.  Data from a mini--scan at the average $\sqrt{s}=4010.4$~MeV ($\langle|Q^2|\rangle=16.08$~GeV$^2$, $\mathcal{L}=20.7$~pb$^{-1}$) and $\sqrt{s}=4260$~MeV ($|Q^2|=18.25$~GeV$^2$, $\mathcal{L}=12.9$~pb$^{-1}$)
have also been analyzed.  


No measurements of the branching fractions for non--$D\overline{D}$ two--body decays of either $\psi(3772)$ or $\psi(4160)$ to light hadrons exist~\cite{pdg}, but they can be estimated using the perturbative QCD (pQCD) prediction that the hadronic decays of $\psi(nS)$ states scale with the principle quantum number $n$ in the same way as the dilepton decays.  The measured values, $\mathcal{B}(\psi(3770)\to e^+e^-)$ and $\mathcal{B}(\psi(4160)\to e^+e^-)$, are $\approx10^3$ times smaller than $\mathcal{B}(\psi(2S)\to e^+e^-)$~\cite{pdg}.  This leads to the estimate that the branching fractions for $\pi^+\pi^-$, $K^+K^-$, and  $p\bar{p}$ decays of $\psi(3770)$ and $\psi(4160)$ are approximately $0.9\times10^{-8}$, $9\times10^{-8}$, and $3.2\times10^{-7}$, respectively, based on the measurements for $\psi(2S)$ decays~\cite{pdg,cleointerf}.  This leads to the estimates that with more than 5.2~million $\psi(3772)$ and $\psi(4160)$ formed in the present measurements, the numbers of resonantly produced  $\pi^+\pi^-$, $K^+K^-$, and $p\bar{p}$ are  $\sim0.04$, 0.4, and 1.8, respectively.  In other words, the resonance contribution to their yield is expected to be vanishingly small, and the counts observed in the present analysis can be entirely attributed to the timelike electromagnetic form factors.  

The event selection and particle identification for the analysis reported in this Letter are similar to those in our earlier publication~\cite{ffcleo}.  The reconstructed events are required to have two charged tracks, zero net charge, and $|\cos\theta|<0.80$.  Each charged particle is required to satisfy the standard criteria for track quality and origin of the track at the interaction point.  In order to develop data-independent particle identification criteria, and to determine event selection efficiencies, Monte Carlo (MC) samples were generated for $e^+e^-\to h^+h^-$, $h=p,\pi,K$ using the EvtGen generator~\cite{evtgen}, and $e^+e^-\to l^+l^-$, $l=e,\mu$, using the Babayaga generator~\cite{babayaga}.  For $e^+e^-\to\pi^+\pi^-$ and $K^+K^-$, the Monte Carlo samples were generated with $\sin^2\theta$ angular distributions, since for pseudoscalar mesons, the differential cross sections are
\begin{equation}
d\sigma_0(s,\theta)_m/d\Omega = (\alpha^2/8s)\beta^3_m |F_m(s)|^2 \sin^2\theta,
\end{equation}
where $\alpha$ is the fine structure constant, $c\beta_m$ is the meson velocity in the laboratory frame, and $F_m(s)$ is the form factor for timelike momentum transfer at $s=|Q^2|$.  For $e^+e^-\to p\bar{p}$ the Monte Carlo samples were generated with both $(1+\cos^2\theta)$ and $\sin^2\theta$ angular distributions, since
\begin{multline}
 d\sigma_0(s,\theta)_p/d\Omega = (\alpha^2/4s)\beta_p [ |G_M(s)|^2(1+\cos^2\theta) \\
 + \tau |G_E(s)|^2\sin^2\theta ],
\end{multline}
where $G_E(s)$ and $G_M(s)$ are the electric and magnetic form factors of the proton, respectively, and $\tau\equiv 4m_p^2/s$.

Monte Carlo generated momentum distributions for different pairs of hadrons ($h^+h^-$) and leptons ($l^+l^-$) show that the peaks for individual particles are broadened by detector resolution, and they overlap and develop large tails at low momenta, mainly due to final state radiation.  Further, the several orders of magnitude larger QED yields of leptons ($e^+e^-$ and $\mu^+\mu^-$) overwhelm the hadron peaks.  The first step of analysis therefore consisted of minimizing the lepton contributions and the radiative tails.  This preselection of events was done by removing events with $E_{CC}$(energy loss in the central calorimeter)/$p$(track momentum)$~>~0.85$ (which reduced $e^+e^-$ by a factor $\sim10^5$), removing those with any hits in the muon chambers (which reduced muons by a factor $\sim10^2$), and requiring that the vector sum of the momenta of the pair be $\sum \vec{p}_{+,-} <60$~MeV/$c$ (which greatly reduced the radiative tails). To develop further event selection criteria, we use the variable $X_h\equiv (E(h^+)+E(h^-))/\sqrt{s}$, as in Ref.~\cite{ffcleo}.

For particle identification in the RICH detector, we define the parameter $L_i$, which is the likelihood that a particle corresponds with a given hypothesis of being of species, $i$, based on the Cherenkov photons detected in the RICH detector.  We combine this with $dE/dx$ information from the drift chambers, $\chi^2(dE/dx)$, defined as  $\chi^2(dE/dx) = [(dE/dx)_\mathrm{measured} - (dE/dx)_\mathrm{expected}]^2/\sigma^2_\mathrm{measured}$, to construct the joint variable $\Delta \mathcal{L}(i,j)$ to distinguish particle type $i$ from contaminant particle type $j$,
\begin{multline}
\Delta \mathcal{L}(i,j) = -2[\log(L_i) - \log(L_j)] \\
 + [ \chi^2(dE/dx)_i - \chi^2(dE/dx)_j ].
\end{multline}



Monte Carlo simulations show that the default values of $\Delta \mathcal{L}(i,j)<0$ are very effective in distinguishing between the desired particle $i$ and the contaminant particle $j$.  Accordingly, we require $\Delta \mathcal{L}(p,\mu)$, $\Delta \mathcal{L}(p,e)$, $\Delta \mathcal{L}(K,\mu)$, $\Delta \mathcal{L}(K,e)$, $\Delta \mathcal{L}(K,\pi)$,  $\Delta \mathcal{L}(K,p)$, $\Delta \mathcal{L}(\pi,e)$, $\Delta \mathcal{L}(\pi,K)$, $\Delta \mathcal{L}(\pi,p)<0$.    Because pions and muons have similar masses, a $\Delta \mathcal{L}(i,j)$ cut is not effective in distinguishing between pions and muons.  It was shown in Ref.~\cite{ffcleo} that $E_{CC}$ can be very effective in distinguishing muons, which suffer only ionization loss in the CC crystals, from pions, which suffer additional energy loss due to hadronic interactions in the crystals. We therefore impose the additional requirement  of $E_{CC}>350$~MeV. With this additional requirement, muon contamination in the pion peak is found to be less than 1\%, and the $X_{\pi,K,p}$ distributions for both $\sqrt{s}=3772$~MeV and 4170~MeV are all free of contaminants, as shown in Fig.~1.

\begin{figure}
\includegraphics[width=3.3in]{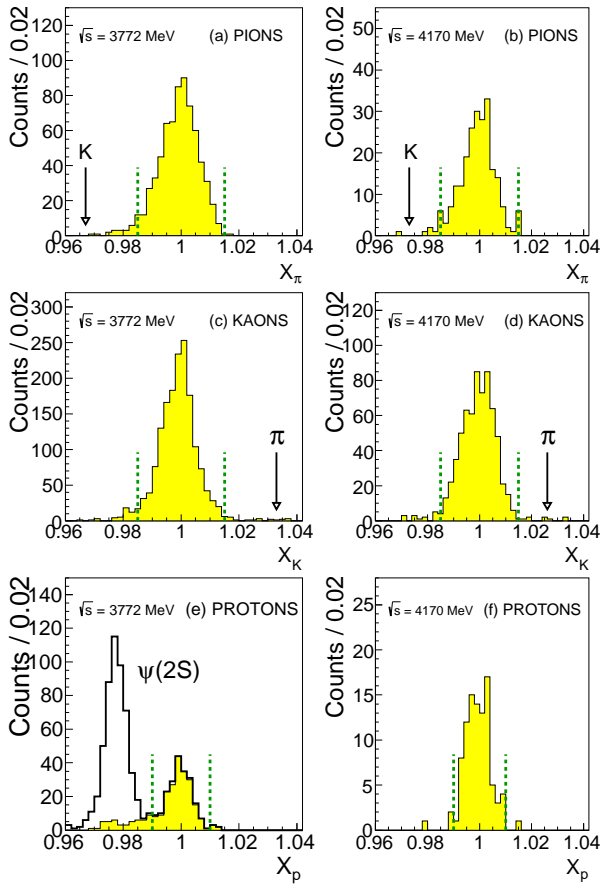}

\caption{Distributions of $X_h\equiv[E(h^+)+E(h^-)]/\sqrt{s}$ for $h\equiv \pi$, $K$, $p$ for $\sqrt{s}=3772$~MeV and 4170~MeV data.
The vertical dashed lines bracket the $X_h$ region in which counts are accepted. 
The open histogram in panel~(e) corresponds to $p\bar{p}$ from $\psi(2S)$ ISR excitation (see text).  Arrows indicate expected positions of contaminants.}
\end{figure}

Since the peaks in Fig.~1 have essentially no backgrounds, we obtain the number of counts $N_\pi(s)$ and $N_K(s)$ as counts in the range $X_{\pi,K}=1.000\pm0.015$, and $N_p(s)$ as counts in the range $X_p=1.000\pm0.010$. We note that the $X_p$ distribution for $\sqrt{s}=3772$~MeV in Fig.~1(e) has a definite tail in the low $X_p$ region. As shown by the unshaded histogram in Fig.~1(e), this arises from $p\bar{p}$ from the decay of $\psi(2S,3686)$ produced by initial state radiation~(ISR), and is clearly observed when the net vector momentum in the events is increased to $\sum\vec{p}<150$~MeV/$c$. 

The observed cross sections are obtained as
$\sigma_0(s) = N(s)/[\epsilon(s)\cdot\mathcal{L}(s) ]$,
where $\epsilon(s)$ is the MC-determined efficiency, and $\mathcal{L}(s)$ is the integrated $e^+e^-$ luminosity.  
The efficiencies at $\sqrt{s}=3772$~MeV and 4170~MeV are respectively 16.2\% and 16.2\% for $\pi^+\pi^-$, 60.2\% and 60.9\% for $K^+K^-$, and 71.3\% and 68.7\% for $p\bar{p}$.
These cross sections are corrected for ISR to obtain the Born cross sections using the method of Bonneau and Martin~\cite{bonneaumartin}.  At $\sqrt{s}=3772$~MeV and 4170~MeV the correction factors are respectively 0.797 and 0.796 for $\pi^+\pi^-$, 0.817 and 0.809 for $K^+K^-$, and 0.806 and 0.800 for $p\bar{p}$.  The Born cross sections are related to the angle integrated cross sections
\begin{flalign}
\sigma_B(s)_{\pi,K}  = (\pi\alpha^2\beta^3_{\pi,K}/3s) |F_{\pi,K}(s)|^2, \hspace*{2.1cm} & \\
\sigma_B(s)_{p}      = (4\pi\alpha^2/3s) \beta_p [ |G_M(s)|^2 + (\tau/2) |G_E(s)|^2 ]. &
\end{flalign}
The cross sections $\sigma_B$ and the form factors $F_{\pi,K}$ and $G_M$ (assuming $G_E=G_M$) derived from them are listed in Table~I.   The systematic uncertainties listed in the table are as described later.  In the Table, and elsewhere in this Letter, the errors indicated in the parentheses are in the corresponding last digits of the main values.  



\begin{figure*}
\includegraphics[width=2.48in]{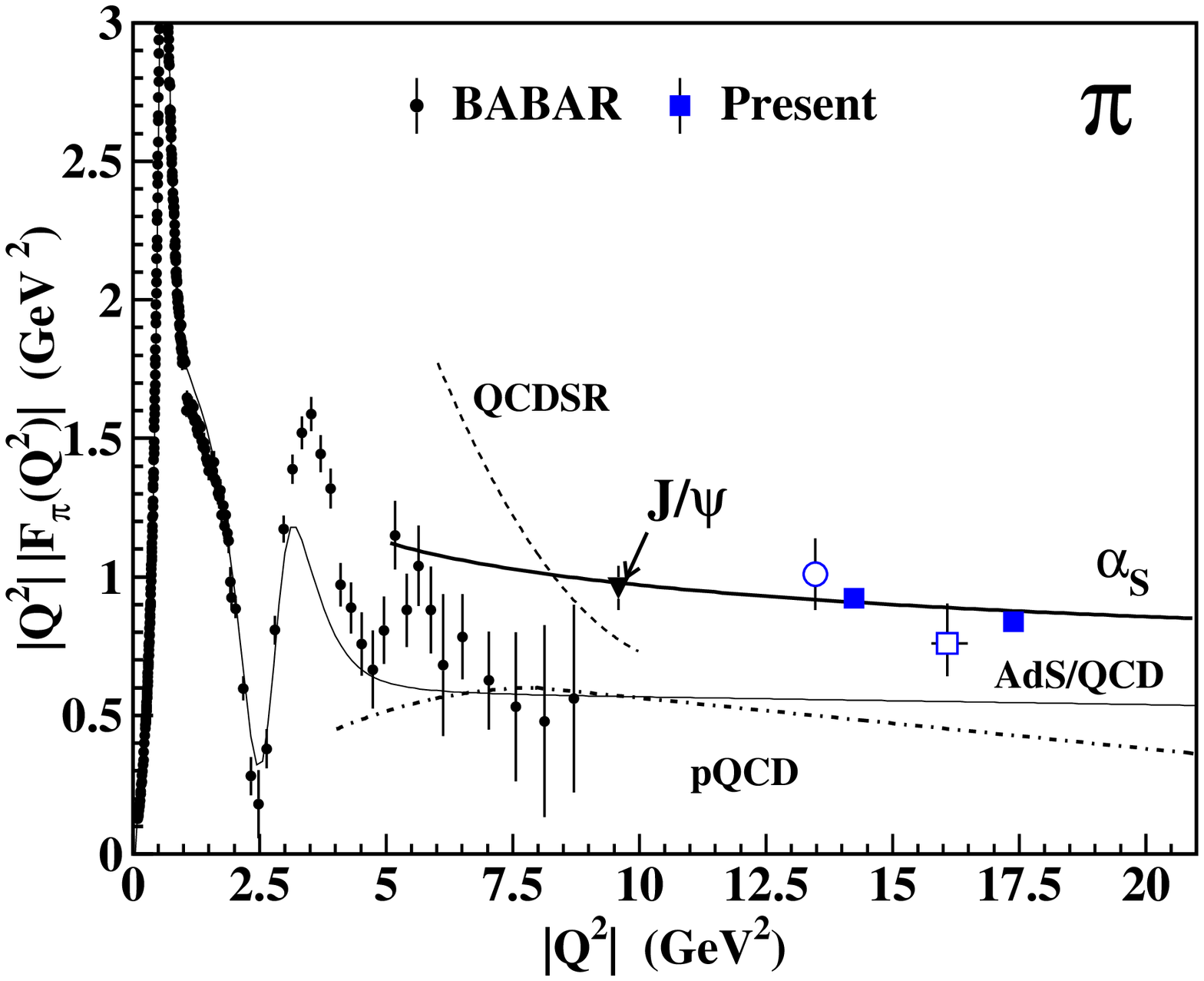}
\hspace*{-20pt}
\includegraphics[width=2.48in]{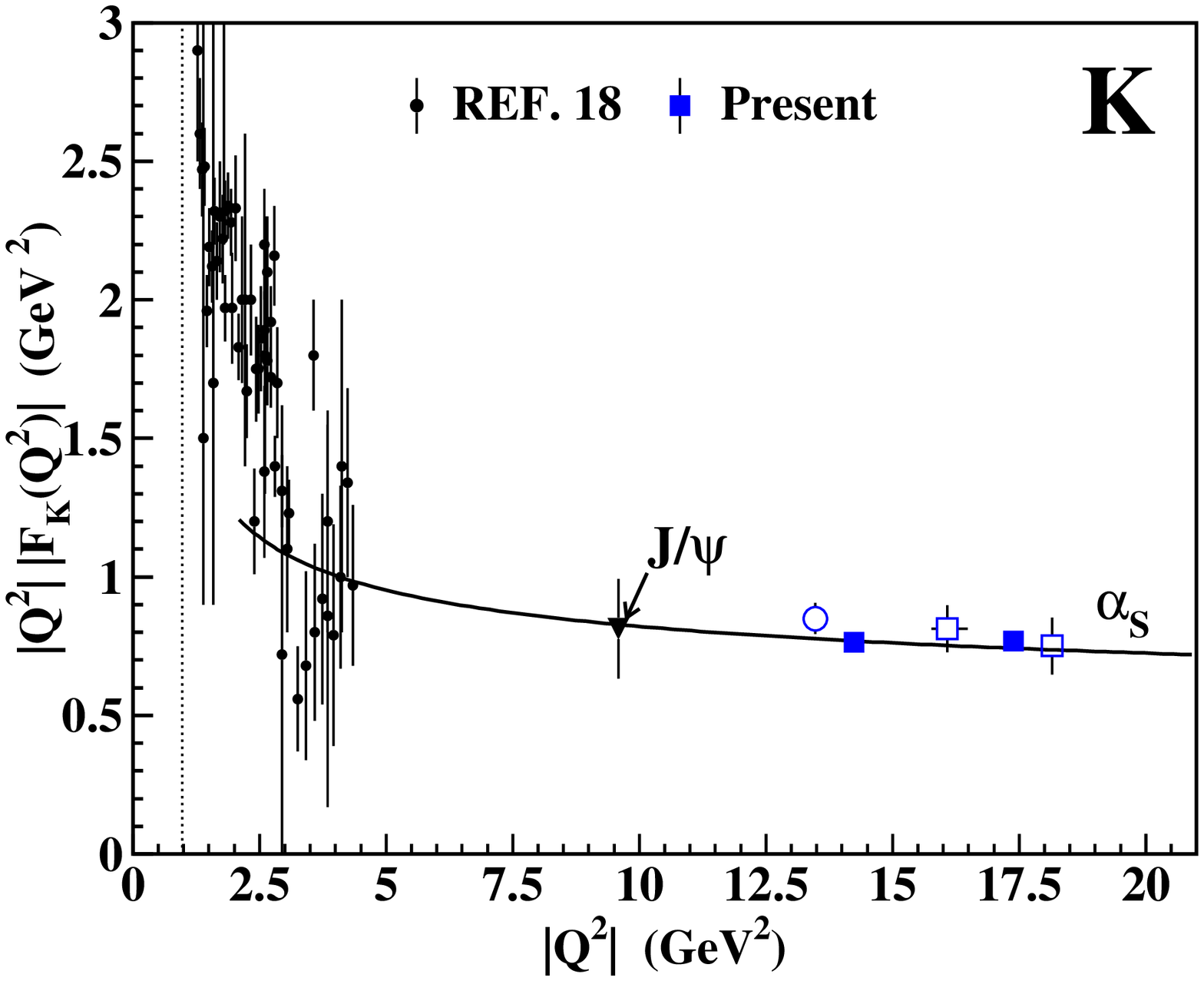}
\hspace*{-20pt}
\includegraphics[width=2.48in]{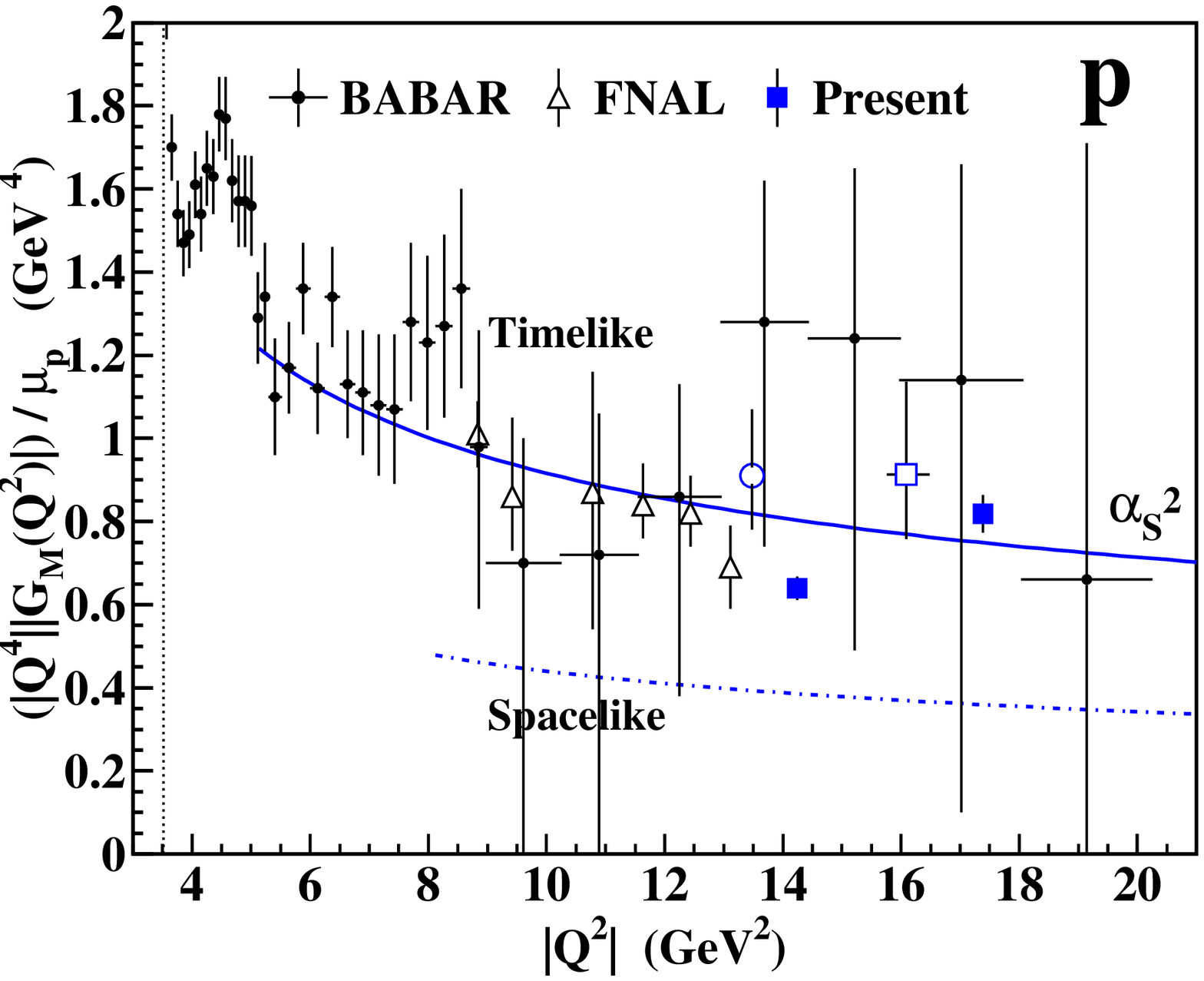}

\caption{Summary of form factor results.   
The solid points in the pion and proton panels are from BaBar ISR measurements~\cite{ffbabar,ffpibabar}, the open triangles are from FNAL $p\bar{p}$ measurements~\cite{ffe760-835}, the open circles are from the CLEO measurements~\cite{ffcleo}, the open squares at $|Q^2|=16.1$~GeV$^2$ and 18.3~GeV$^2$ are from our small statistics data sets, and the solid squares are from the present measurements at $|Q^2|=14.2$~GeV$^2$ and 17.4~GeV$^2$. 
The theoretical curves for pions are from References~[16,17,18]. 
The solid curves illustrate the arbitrarily normalized variation of $\alpha_S$ for $\pi$ and $K$, and $\alpha_S^2$ variation for protons. }
\end{figure*}

\begin{table}
\caption{Cross sections for $e^+e^-\to\pi^+\pi^-$, $K^+K^-$, and $p\bar{p}$ for $e^+e^-$ annihilations at $\sqrt{s}=3772$~MeV and 4170~MeV, and the corresponding form factors of pion, kaon, and proton.}
\begin{ruledtabular}
\begin{tabular}{lcccc}
$\pi^+\pi^-,K^+K^-$ & $N_{\pi,K}$ & $\sigma_B$ (pb) & $10F_{\pi,K}$ & $|Q^2|F_{\pi,K}$  \\
 $p\bar{p}$    & $N_p$       & $\sigma_B$ (pb) & $10^2G_M$   & $|Q^4|G_M/\mu_p$ \\
\hline
\multicolumn{5}{l}{ $\sqrt{s}=3772$~MeV, $|Q^2|=14.2$~GeV$^2$} \\
\hline
$\pi^+\pi^-$ &  661(26)  & 6.36(25)(36) & 0.65(1)(2) & 0.92(2)(3)  \\
$K^+K^-$     & 1564(40)  & 3.95(10)(22) & 0.54(1)(1) & 0.76(1)(2)  \\
$p\bar{p}$   &  213(15)  & 0.46(3)(3)  & 0.88(3)(2) & 0.64(2)(2)  \\
\hline
\multicolumn{5}{l}{ $\sqrt{s}=4170$~MeV, $|Q^2|=17.4$~GeV$^2$} \\
\hline
$\pi^+\pi^-$ &  218(12) & 2.89(16)(16) & 0.48(1)(1) & 0.84(2)(2)  \\
$K^+K^-$     &  644(25) & 2.23(9)(12)  & 0.44(1)(1) & 0.77(2)(2)  \\
$p\bar{p}$   &   92(10) & 0.29(3)(2)  & 0.76(4)(2) & 0.82(4)(2)  \\

\end{tabular}
\end{ruledtabular}
\end{table}


\noindent \textbf{Pion Form Factors}---Timelike form factors of pions, measured by $e^+e^-\to\pi^+\pi^-$, have been reported by Babar for $0.09~\mathrm{GeV}^2\le|Q^2|\le8.7~\mathrm{GeV}^2$ by the ISR method~\cite{ffpibabar}.  At the largest $|Q^2|$, these results have up to $\pm100\%$ errors, and were based on $<10$~counts/bin.
Our earlier measurement~\cite{ffcleo} at $|Q^2|=13.48$~GeV$^2$ had only $26(5)$ observed counts, and resulted in $F_\pi(13.48~\mathrm{GeV}^2)=0.075(9)$.  The present results, $F_\pi(14.2~\mathrm{GeV}^2)=0.065(2)$ and $F_\pi(17.4~\mathrm{GeV}^2)=0.048(1)$ are based on $661(26)$ and $218(12)$ observed counts, respectively.  These are listed in Table~I.  
In Fig.~2 we plot our results together with all previous results, including the indirect result for $|Q^2|=M^2(J/\psi)$~\cite{milana}.    As shown in the figure, all measurements with $|Q^2|>9$~GeV$^2$ are in excellent agreement with the dimensional counting rule prediction of a $1/|Q^2|$ variation of the form factors at large momentum transfers.  Also shown in the figure are three illustrative theoretical predictions.  
 The $Q^2$ behavior of the prediction of the QCD sum rule inspired model~\cite{bakulev}  disagrees strongly with the data.  The pQCD prediction of Gousset and Pire~\cite{goussetpire} is nearly a factor two smaller than our measurements, and the latest AdS/QCD prediction by Brodsky and de~Teramond~\cite{guybrodsky} reproduces the data below 5~GeV$^2$, but falls to 2/3 of the observed values for $|Q^2|>5$~GeV$^2$.  Czyz et al.~\cite{czyz} have shown that the measured $F_\pi(|Q^2|)$ at $|Q^2|>5$~GeV$^2$ can be parameterized in a VDM approach, but only by including hypothesized radial $\rho_3$, $\rho_4$, $\rho_5$ resonances.


%

\textbf{Kaon Form Factors}---The results of our present measurements at $|Q^2|=14.2$~GeV$^2$ and 17.4~GeV$^2$ are listed in Table~I.  
In Fig.~2, we show our present results for kaon form factors along with early results for $|Q^2|<4.4$~GeV$^2$~\cite{ffkk}, the indirect result for $|Q^2|=M^2(J/\psi)$~\cite{seth}, and our previous measurement at $|Q^2|=13.48$~GeV$^2$~\cite{ffcleo}.  As for pions, all results for $|Q^2|>9$~GeV$^2$ follow the predicted $\alpha_S/|Q^2|$ behavior of the form factors.  No theoretical predictions for kaon timelike form factors exist.  An empirical fit to the data has however been made by Czyz et al.~\cite{czyz}, but it requires an ``infinite tower'' of hypothetical $\rho$, $\omega$, and $\phi$ resonances.

\noindent \textbf{Proton Form Factors}---In Fig.~2, we show the previously published proton form factors for timelike momentum transfers. 
At the largest $|Q^2|$ all these measurements were severely limited by statistics, with 0 and 1 counts at $|Q^2|=14.4$~GeV$^2$~\cite{ffe760-835},  16(5) counts at $|Q^2|=13.48$~GeV$^2$~\cite{ffcleo}, and  8(5) counts in the region $|Q^2|=13.0-14.4$~GeV$^2$~\cite{ffbabar}.  We observe 213(15) and 92(10) counts at $|Q^2|=14.2$~GeV$^2$ and 17.4~GeV$^2$, respectively.  This allows us to examine, for the first time, the $|Q^2|$ dependence of the proton form factors sensitively.
As listed in Table~I, and shown in Fig.~2, we find that in disagreement with the dimensional counting rule prediction of the weak $\alpha_S^2$ variation of $|Q^4|G_M(|Q^2|)/\mu_p$, its value 0.64(3)~GeV$^4$ at $|Q^2|=14.2$~GeV$^2$ is 22(4)\% smaller than 0.82(5)~GeV$^4$ at 17.4~GeV$^2$. We note, however, that $|Q^2|G_M(|Q^2|)/\mu_p$ is essentially constant, being $=4.5(2)\times10^{-2}$~GeV$^2$ and $4.7(2)\times10^{-2}$~GeV$^2$ for $|Q^2|=14.2$~GeV$^2$ and 17.4~GeV$^2$, respectively.  This is reminiscent of the fact that for the spacelike momentum transfers for the equivalent form factors, $F_1(\mathrm{Dirac})$ and $F_2(\mathrm{Pauli})$, it is $QF_2/F_1$ which is found to be constant, rather than $Q^2F_2/F1$ as predicted~\cite{qcd}.

Recent polarization measurements of the spacelike form factors of proton at JLab have revealed that $G_E(Q^2)/G_M(Q^2)$ monotonically decreases with increasing $Q^2$~\cite{jlab-proton}.  
We fit the summed differential cross sections we measure at $\sqrt{s}=3772$~MeV and 4170~MeV with Eq.~2, and obtain $G_M=(0.85\pm0.07)\times10^{-2}$, $G_E=(0.71\pm1.17)\times10^{-2}$, and $G_E/G_M=0.83^{+0.98}_{-0.67}$ at the average $\langle|Q^2|\rangle=16.1$~GeV$^2$.  This shows that we have little sensitivity to $G_E$, and even with our larger statistics we can not determine $G_E/G_M$.  The results listed in Table~1 and shown in Fig.~2 are assuming $G_E(|Q^2|)=G_M(|Q^2|)$.
If $G_E(|Q^2|)=0$ is assumed, $G_M(|Q^2|)$ would increase by $\sim6\%$ at $|Q^2|=14.2$~GeV$^2$, and by $\sim5\%$ at  $|Q^2|=17.4$~GeV$^2$.


%

\noindent \textbf{Systematic Uncertainties}---As in our previous publication~\cite{ffcleo}, we estimate uncertainties of 1\% in trigger efficiencies, 2\% in tracking efficiencies, 1\% in luminosity, and 0.2\% in radiative corrections.  In addition, we estimate the total uncertainty due to variation of $\sum\vec{p}$, $E_{CC}$, and $\Delta \mathcal{L}$ to be $<5\%$ for $\pi$, $K$, and $p$ at both $\sqrt{s}=3772$~MeV and 4170~MeV.  This brings the total systematic uncertainty to 6\%.

To summarize, we have made form factor measurements for pions, kaons, and protons for the highest timelike momentum transfers of $|Q^2|=14.2$~GeV$^2$ and 17.4~GeV$^2$ with nearly five times higher precision than prior measurements in the literature.
For pions and kaons we find that the dimensional counting rule prediction of $\alpha_S/Q^2$ variation of the form factors with $|Q^2|$ holds very well.  However, the existing theoretical predictions for pions fail by large factors to predict the magnitude of the form factors. We find $F_\pi(|Q^2|)/F_K(|Q^2|)=1.21(3)$ and 1.09(4) for $|Q^2|=14.2$~GeV$^2$ and 17.4~GeV$^2$, respectively.  These are in agreement with $F_\pi(|Q^2|)/F_K(|Q^2|)=1.19(17)$ measured at 13.48~GeV$^2$~\cite{ffcleo}, and in large disagreement with the asymptotic prediction that they should be equal to the ratio of the squares of the decay constants, $f^2_\pi/f^2_K=0.67(1)$.
This might be indicative of the difference between the distribution functions of pion and kaon~\cite{qcd}.

For protons, we find that the timelike form factors continue to be a factor two or more larger than the corresponding spacelike form factors, as shown in Fig.~2.  We find the unexpected result that $|Q^4|G_M(|Q^2|)/\mu_p$ at $|Q^2|=14.2$~GeV$^2$ is 22(4)\% smaller than at $|Q^2|=17.4$~GeV$^2$.  The difference is suggestive of the near--constancy of $|Q^2|G_M(|Q^2|)/\mu_p$, instead.  


The overall conclusion of the present investigation is that the asymptotic predictions of QCD-based models are not realized even at momentum transfer as large as 18~GeV$^2$, and that theoretical understanding of timelike form factors of hadrons is still lacking at the quantitative level, and our precision measurements provide new challenges for theory.

This investigation was done using CLEO data, and as members of the former CLEO Collaboration we thank it for this privilege. 
This research was supported by the U.S. Department of Energy.

\end{document}